\documentclass[reprint,amsmath,amssymb,aps,prb,superscriptaddress,twocolumn]{revtex4-2}
\usepackage{graphicx}
\usepackage{graphics}
\usepackage{dcolumn}
\usepackage{bm}
\usepackage{amsfonts}
\usepackage{amssymb}
\usepackage{xcolor}
\usepackage{multirow}
\usepackage{mathtools}
\usepackage[tight]{subfigure}
\usepackage{relsize}

\definecolor{green}{rgb}{0,0.6,0.1}

\begin{document}

\title{{\it Ab initio} study of orbital-selective superconductivity in $\gamma$-BiPd}

\author{Sonu Prasad \surname{Keshri}}
\affiliation{Department of Physics, National Taiwan University, Taipei 10617, Taiwan\looseness=-1}

\author{Guang-Yu Guo}
\email{gyguo@phys.ntu.edu.tw}
\affiliation{Department of Physics, National Taiwan University, Taipei 10617, Taiwan\looseness=-1}
\affiliation{Physics Division, National Center for Theoretical Sciences, Taipei 10617, Taiwan\looseness=-1}

\date{\today}

\begin{abstract}
We investigate the superconducting (SC) properties of experimentally realised $\gamma$-BiPd
by solving the anisotropic Migdal-Eliashberg equations in conjunction with {\it ab initio} relativistic
calculations of the electron and phonon band structures as well as electron-phonon coupling (EPC) matrix elements.
Our study reveals that $\gamma$-BiPd possesses a complex Fermi surface (FS), consisting of two electron pockets and one hole pocket, 
each characterised by distinct atomic orbitals. Our key finding is that the superconductivity in 
$\gamma$-BiPd is primarily orbital-selective, arising from Bi $p$-orbitals, and distributed anisotropically on the FS,
although contribution from Pd $d$-orbitals, particularly on the hole pocket, is also discernable. 
While our results show an anisotropic nature of the {\bf k}-dependent SC gap $\Delta_{\bf k}$ and EPC strength $\lambda_{\bf k}$
across the FS, calculated superconducting quasiparticle density of states $N_S$ spectra exhibit a U-shaped gap 
and $\Delta_{\bf k}$ distribution forms a single peak, being consistent with the spin-singlet $s$-wave superconductivity 
observed in this material.
The calculated $T_c$ is $\sim$2.0 K, agreeing in order of magnitude with the experimental value of 3.3 K 
in $\gamma$-BiPd thin films.
The predicted EPC-enhanced Sommerfeld coefficient $\gamma_n$ of $0.141$ mJ/K$^2$cm$^3$
is similar to the experimental $\gamma_n$ value ($0.119$ mJ/K$^2$cm$^3$) of the isoelectronic
and isostructural Bi(Pd$_{0.5}$Pt$_{0.5}$) alloy.
\end{abstract}

\maketitle
\section{INTRODUCTION}

The discovery of strong $\bf{k}$-dependent and multigap superconductivity, which was observed experimentally 
as early as 1980~\cite{Two-Band_Binnig1980} and extensively studied theoretically~\cite{Bardeen-Cooper_Suhl1959} 
since the advent of the Bardeen–Cooper–Schrieffer (BCS) theory~\cite{Microscopic_BCS1957}, 
has provided new physical insights. These include chiral ground states~\cite{Two_Komendova2012}, 
vortices with fractional flux~\cite{Vortices_Babaev2002}, and topological solitons~\cite{Topological_Garaud2011}. 
Anisotropic and multigap superconductivity can be introduced by the combined effects of the normal state quantities
$N_{F}$, $V$, and $\omega_c$~\cite{Two-Band_Binnig1980}, which enter in the superconducting (SC) 
gap $\Delta$ in the weak-coupling isotropic BCS limit as $\Delta=2\hbar\omega_c e^{[-2/N_{F}V]}$, 
where $N_{F}$ is the density of states (DOS) near the Fermi level ($E_F$), $V$ the electron-phonon coupling (EPC) potential, 
and $\omega_c$ the cutoff phonon angular frequency~\cite{Microscopic_BCS1957,Room_Pickett2023}.
Another significant quantity from the normal state is the Fermi velocity (${v_f}$),
which plays a role in the current density for penetration phenomena.

In systems with multiple bands crossing the $E_F$, each band can contribute differently 
to $N_{F}$ and $V$, depending on the correlation strength of each band or orbital~\cite{Bardeen-Cooper_Suhl1959}.
Binary compound MgB$_2$ has been shown multiband ($\sigma$ and $\pi$ bands) superconductivity 
with transition temperature $T_c\sim39$ K where EPC is the sole 
source of pairing~\cite{Superconductivity_Nagamatsu2001,Complex_Hinks2001,Specific_Bouquet2001,
Two-Band_Iavarone2002,Origin_Souma2003}. Subsequently many Fe-based
superconductors such as FeSe~\cite{Discovery_sprau2017,Orbital_nica2017,Orbital_Kreisel2017,
Imaging_kostin2018,Orbital_hu2018}, LaFeAsO$_{1-x}$F$_x$-based 
superconductors~\cite{Unconventional_Kuroki2008,Unconventional_Mazin2008,Even_Dai2008,Role_yi2017,
Multiorbital_nica2021}, and many others including NbS$_2$~\cite{Origin_Heil2017,Orbital_Bi2022,
Superconducting_Guillamon2008}, NbSe$_2$~\cite{Fermi_yokoya2001}, and 
TiTe$_2$~\cite{Orbital_antonelli2022} have been shown multiple electron and hole pockets. In FeSe, 
Fe $d_{xz}$, $d_{yz}$ and $d_{xy}$ orbitals dominate the bands near the $E_F$ and thus would lead to 
the orbital-selective superconductivity. The interplay
of signs of electron and hole pockets on the Fermi surface (FS) has led to the classification
of $s$-wave superconductors as s$_{\pm}$ and $s_{++}$ waves~\cite{Unconventional_Kuroki2008,Sudden_tafti2013}.

Bi-Pd-based alloys have recently attracted significant interest in the scientific community 
due to their potential for topological superconductivity and the discovery of various pairing 
mechanisms~\cite{Superconductivity_Joshi2011,Superconductivity_imai2012,
Topologically_sakano2015,Dirac_sun2015,Full_iwaya2017,Single_Kacmarcik2016,Probing_Mitra2017,
Unequivocal_Chiang2023,Sharma2024}. 
These alloys exist in multiple allotropes and phases~\cite{Unequivocal_Chiang2023,Full_heise2014}.   
Among them, $\alpha$-BiPd, $\beta$-Bi$_2$Pd and $\gamma$-BiPd were recently found to show 
$T_c$ of 3.7, 3.6, and 3.3 K, respectively~\cite{Unequivocal_Chiang2023}. 
Tetragonal $\beta$-Bi$_2$Pd (I$_4$/mmm) and hexagonal $\gamma$-BiPd (P6$_3$/mmc) are centrosymmetric, 
and were reported to show triplet pairing and singlet pairing, respectively~\cite{Unequivocal_Chiang2023}. 
The centrosymmetric $\gamma$-BiPd would transform to noncentrosymmetric $\alpha$-BiPd 
after post-annealing at 270$^0$ C. Thus, monoclinic $\alpha$-BiPd (P2$_1$)~\cite{Superconductivity_Joshi2011} 
would exhibit an admixture of singlet and triplet pairing~\cite{Unequivocal_Chiang2023,Spin_Xu2020} 
because of the lack of inversion symmetry combined with Rashba spin-orbit coupling 
(SOC)~\cite{Superconductivity_sigrist2007,Superconductivity_kneidinger2015,Spin_Xu2020}.
In $\alpha$-BiPd, topological phenomena such as Dirac surface states have been observed, 
and the material has been investigated for topological superconductivity~\cite{Dirac_sun2015,Unusual_Thirupathaiah2016}. 
The SOC would also introduce exotic proximity effects at
superconductor/ferromagnet (S/F) interface~\cite{Proximity_lupke2020,Tuning_Ben2010,Tunable_Caviglia2010,
Coexistence_Dikin2011,Superconducting_Michaeli2012,Spin_flokstra2023}. In this context, SOC acts as an exchange interaction, 
converting spin-singlet pairing into spin-triplet pairing.
It also gives rise to a new class of superconductors known as type I and type II Ising 
superconductors~\cite{Superconductivity_saito2016,Evidence_lu2015,Ising_xi2016,falson2020type}.

Despite the different crystal structures and pairing mechanisms, the three Bi-Pd alloys  
have almost the same $T_c$. Similar to $\alpha$-BiPd, $\beta$-Bi$_2$Pd is also topologically nontrivial,
and its SC properties are extensively studied in the quest for novel phenomena such as Andreev bound states 
and Majorana fermions~\cite{Superconductivity_imai2012,Topologically_sakano2015,Full_iwaya2017,
Single_Kacmarcik2016,Unequivocal_Chiang2023}. The upper critical magnetic field and specific heat measurements 
initially suggested that $\beta$-Bi$_2$Pd could be a multigap superconductor~\cite{Superconductivity_imai2012}.
However, subsequent investigations using scanning tunneling microscopy and muon-spin relaxation experiments 
confirmed that $\beta$-Bi$_2$Pd is actually a single-gap BCS superconductor~\cite{Single_Kacmarcik2016}.

\begin{figure}[!htb]
\centering
\includegraphics[width=0.80\columnwidth]{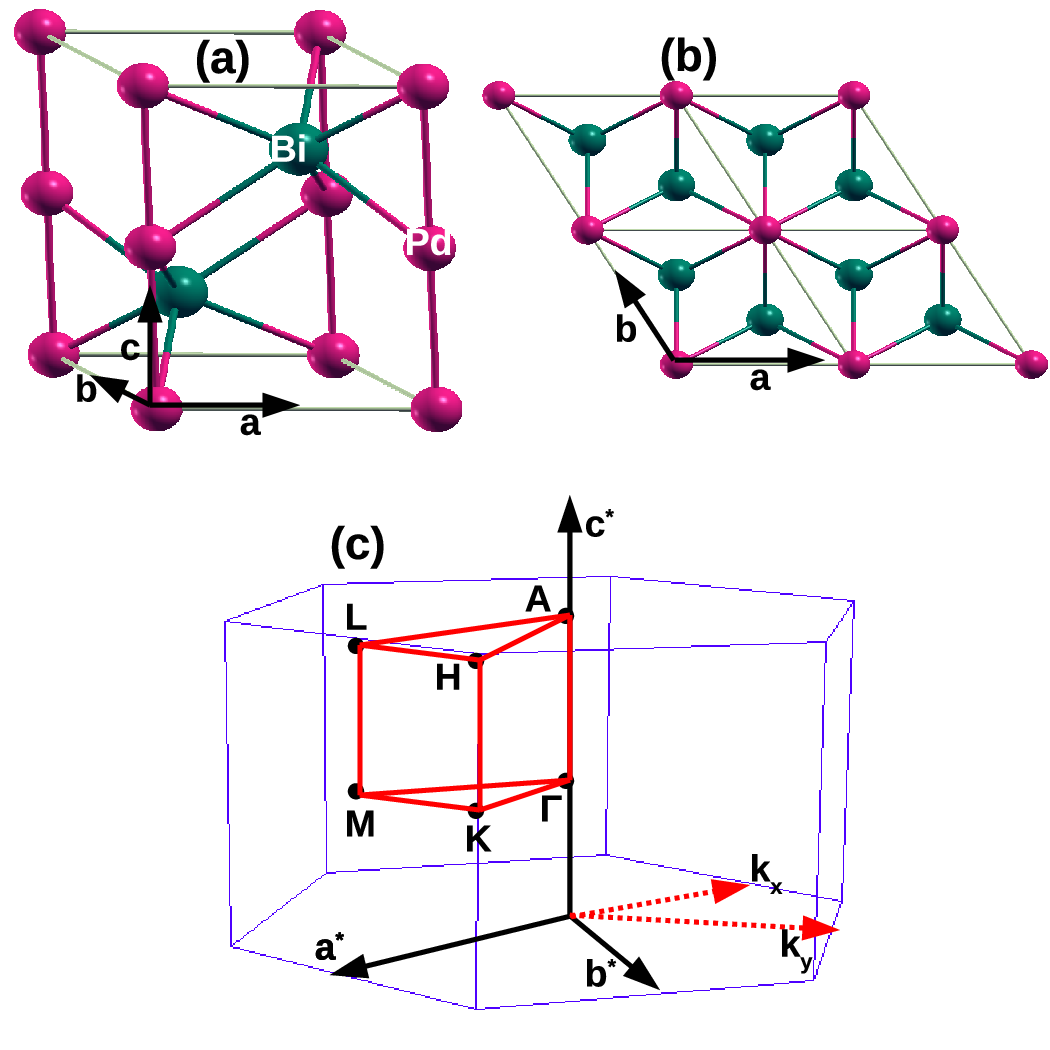}
\caption{(a) A side-view and (b) the $z$-axis view of crystal structure of $\gamma$-BiPd.
In (b), the 2$\times$2$\times$1 supercell is shown. The associated BZ is illustrated in (c).}
\label{fig:crystal}
\end{figure}

Among the three Bi-Pd alloys, $\gamma$-BiPd has received relatively little theoretical attention despite 
its nontrivial topology~\cite{Unequivocal_Chiang2023,Sharma2024}.
$\gamma$-BiPd features a simple hexagonal crystal structure (Fig. \ref{fig:crystal}) and exhibits 
spin-singlet $s$-wave superconductivity~\cite{Unequivocal_Chiang2023}. 
Yet, its $T_c$ matches closely with those of $\alpha$-BiPd and $\beta$-Bi$_2$Pd, which raises intriguing questions 
about the microscopic normal state features that lead to these distinct SC states.
To explore these questions, we have investigated the SC properties of $\gamma$-BiPd 
by solving the anisotropic Migdal-Eliashberg equations and then evaluating 
${\bf{k}}$-dependent SC gap $\Delta_{\bf k}$, $T_c$ and SC quasiparticle DOS ($N_s$). 
Our primary finding is that the superconductivity in $\gamma$-BiPd is mainly orbital-selective and anisotropic. 
Our calculated $T_c$ is $\sim$2.0 K, being in reasonable agreement with the experimental value of $\sim$3.3 K 
in $\gamma$-BiPd thin films~\cite{Unequivocal_Chiang2023}. The $\Delta_{\bf k}$ on the FS forms a single gap, 
although it is anisotropic, and the $N_s$ spectra exhibit a U-shaped gap. All these findings are consistent 
with the spin-singlet $s$-wave superconductivity observed in $\gamma$-BiPd films~\cite{Unequivocal_Chiang2023}.

The rest of this paper is arranged as follows. In Sec.~\ref{sec:methodology}, we provide an overview of the
{\it ab initio} Migdal-Eliashberg theory of superconductivity adopted in this study.
This section also gives details of various intermediate quantities involved. 
In Sec.~\ref{sec:Computational_details}, we provide the computational details 
of the {\it ab initio} calculations conducted in this study.
In Sec.~\ref{sec:Crystal_Structure}, we report the theoretical crystal structure and calculated phonon dispersion.
In Sec.~\ref{sec:Electronic_Structure}, the electronic structure, FS, and the atomic orbital contributions
on the FS are presented.
In Sec.~\ref{sec:EPC}, we present the EPC strength as a function of phonon energy in the
Brillouin zone (BZ), and examine its distribution on the FS. 
In Sec.~\ref{sec:Superconducting_Properties}, SC properties are presented, including the ${\bf{k}}$-dependent SC 
gap and its orbital characters, the $T_c$, and the $N_s$. 
We summarize the main results of this study in Sec.~\ref{sec:Conclusion}.
Finally, we show the significant SOC effect on the energy bands in the vicinity of the Fermi level 
in $\gamma$-BiPd in Appendix~\ref{sec:appenix}, and also report the calculated Coulomb pseudopotential dependence 
of $T_c$ and SC gap function in Appendix~\ref{sec:appenix_B}.

\section{Methodology}
\label{sec:methodology}
We calculate the ${\bf{k}}$-dependent SC gap, the $T_c$ and SC quasiparticle DOS by solving the anisotropic Migdal-Eliashberg (ME) equations. 
A small energy window near the $E_F$ defined by $\hbar\omega_{ph}$ is important, as
the pair formation takes place in this energy window.
In the vicinity of the FS, there are two anisotropic ME equations to solve, one for the
renormalization function $Z(j{\bf k},i\omega_{n})$ and the other for the order parameter
$\phi({j\bf k},i\omega_{n})=Z({j\bf k},i\omega_{n})\Delta({j\bf k},i\omega_{n})$ given by \cite{Anisotropic_Margine2013}
\begin{align}
Z({j\bf k},i\omega_{n}) = & 1+\frac{\pi T}{\omega_{n}}\mathlarger{\sum}_{{j^{'}\bf{k^{'}}}n^{'}}W_{j^{'}k^{'}}
\frac{\omega_{n}}{\sqrt{R({j^{'}\bf{k^{'}}},i\omega_{n^{'}})}}\nonumber\\&\times\lambda\Big({j\bf{k}},{j^{'}\bf{k^{'}}},n-n^{'}\Big),
\label{eq:1}
\end{align}
and
\begin{align}
Z({j\bf{k}},i\omega_{n}) \Delta({j\bf{k}},i\omega_{n})=& \pi T\mathlarger{\sum}_{{j^{'}\bf{k^{'}}}n^{'}}W_{k^{'}}
\frac{\Delta({j^{'}\bf{k^{'}}},i\omega_{n^{'}})}{\sqrt{R({j^{'}\bf{k^{'}}},i\omega_{n^{'}})}}\nonumber\\&\times\Big[\lambda ({j\bf{k}},{j^{'}\bf{k^{'}}},n-n^{'})-\nonumber\\&N_{F}V(j{\bf{k}}-j^{'}\bf{k^{'}})\Big],
\label{eq:2}
\end{align}
with 
$R({j\bf{k}},i\omega_{n})=\omega^{2}_{n}+\Delta^{2}({j\bf{k}},i\omega_{n})$ and
$W_{{j\bf{k}}}=\delta(\epsilon_{{j\bf{k}}})/N_{F}$, where $i\omega_{n}=i(2n+1)\pi T$ is
fermion Matsubara frequencies, $n$ is an integer and $T$ is the absolute temperature. 
We set $\hbar=k_B=1$.
$V(j{\bf{k}}-j^{'}\bf{k^{'}})$ is the matrix
elements of the static screened Coulomb interaction between
the electronic states ${j\bf{k}}$ and ${j^{'}\bf{k}^{'}}$, and
$\lambda ({j\bf{k}},{j^{'}\bf{k^{'}}},n-n^{'})$ is the anisotropic EPC parameter given by
\begin{align}
\lambda ({j\bf{k}},{j^{'}\bf{k^{'}}},n-n^{'})= &\int_{0}^{\infty}d\omega \frac{2\omega}
{(\omega_{n}-\omega_{n^{'}})^{2}+\omega^{2}}\nonumber\\&\times\alpha^{2}F({j\bf{k}},{j^{'}\bf{k^{'}}},\omega).
\end{align}
The quantity $\alpha^{2}F({j\bf{k}},{j^{'}\bf{k^{'}}},\omega)$ is the anisotropic
Eliashberg spectral function expressed as
\begin{align}
\alpha^{2}F({j\bf{k}},{j^{'}\bf{k^{'}}},\omega)= N_{F}\mathlarger{\sum}_{\nu}|g_{{jj^{'}\bf{kk^{'}}}\nu}|^{2}\delta(\omega-\omega_{{\bf{jk-j^{'}k^{'}}},\nu}).
\end{align}
The electron-phonon matrix elements $g_{{jj^{'}\bf{kk^{'}}}\nu}$ are expressed in terms of
derivatives of the self-consistent potential $\partial_{{\bf{q}}\nu}V_{sc}$ as
$g_{{jj^{'}\bf{kk^{'}}}\nu}=\left\langle \psi_{{j^{'}\bf{k^{'}}}}|\partial_{{\bf{q}}\nu}V_{sc}| \psi_{{j\bf{k}}}\right\rangle$, associated with
a phonon of wave vector ${\bf{q}}$ and branch $\nu$. 
$\psi_{{j\bf{k}}}$ is the electronic wavefunction for wavevector ${\bf{k}}$ in band
$j$. 
The electronic eigen energies $\epsilon_{{j\bf{k}}}$ 
are taken near the $E_F$. 
In the present calculations, the ${\bf{k}}$- and ${\bf{q}}$-meshes are choosen 
to be uniform and they satisfy the condition
${\bf{q}={\bf k}^{'}-{\bf k}}$ and must be commensurate.

The phonon-mode resolved EPC $\lambda_{{\bf{q}}\nu}$ is given by
\begin{align}
\lambda_{{\bf{q}}\nu} = \frac{1}{N_{F}\omega_{{\bf{q}}\nu}}\sum_{{\bf{jk,j^{'}k^{'}}}}W_{{j\bf{k}}}|g_{{jj^{'}\bf{kk^{'}}}\nu}|^{2}\delta(\epsilon_{j\bf{k}})\delta(\epsilon_{j^{'}\bf{k^{'}}}),
\end{align}
where $\delta$ is the Dirac delta function.

The SC quasiparticle DOS relative to the normal-state DOS at the $E_F$ is
\begin{align}
\frac{N_s(\omega)}{N_F}=\mathlarger{\mathlarger{\sum_{j}\int_{BZ}}}\frac{d{\bf k}}{\Omega_{BZ}}\frac{\delta(\epsilon_{j{\bf k}}-\epsilon_{F})}{N_F}\mathrm{Re\Bigg[\frac{\omega}{\sqrt{\omega^{2}-\Delta_{j{\bf k}}^{2}(\omega)}}\Bigg]},
\label{eq:3}
\end{align}
where $\Omega_{BZ}$ is the BZ volume, and $\Delta_{j{\bf k}}$ is the complex
energy-dependent gap for band $j$ and momentum ${\bf k}$.

\section{Computational details}
\label{sec:Computational_details}
The QUANTUM ESPRESSO package \cite{Quantum_giannozzi2009,Advanced_giannozzi2017} is adopted 
here for the structural optimization, electronic structure, and phonon dispersion calculations 
in a fully first-principles way. Because $\gamma$-BiPd contains heavy Bi atoms, 
the presence of relativistic SOC would significantly modify the electronic band structure of $\gamma$-BiPd, 
especially the energy bands near the $E_F$, as demonstrated in Appendix A. Therefore, all the results presented 
in this paper are obtained from the first-principles calculations with the SOC included. 
The Perdew-Burke-Ernzerhof solid (PBEsol) type exchange-correlation functional along with the
optimized norm-conservinng Vanderbilt relativistic pseudopotentials \cite{Optimized_Hamann2013} are used. 
The kinetic-energy cutoff of 100 Ry, and the Methfessel-Paxton smearing width of 0.02 Ry is used. 
The self-consistent field (SCF) charge density is computed on the $\Gamma$-centered $\bf{k}$-mesh of 
12$\times$12$\times$8. For the DOS and FS calculations, a dense $\bf{k}$-mesh of 36$\times$36$\times$18 is used. 
The phonon dispersion and linear variation of the self-consistent potential
were calculated using density functional perturbation theory (DFPT)~\cite{Phonons_Baroni2001} 
on the 6$\times$6$\times$3 $\bf{q}$-mesh with threshold for self-consistency of $10^{-14}$. 
We use the EPW code to calculate the SC properties in the anisotropic ME formalism with the electron-phonon 
interpolation in Wannier manifold \cite{EPW_ponce2016,Anisotropic_Margine2013,Electron_lee2023,EPW_noffsinger2010}. 
Solutions of the ME equations are sensitive to the sampling of electron-phonon matrix elements 
near the $E_F$. To obtain converged results, it requires a dense ${\bf{k}}$ and ${\bf{q}}$-point meshes. 
These shortcomings are overcome by employing interpolation method based on first-principles calculations 
in conjunction with maximally localized Wannier functions \cite{Anisotropic_Margine2013} as implemented 
in the EPW code. The band structure calculation required for the Wannier interpolation 
is done on uniform $\bf{k}$-point mesh of 12$\times$12$\times$6. The ME equations are solved 
by calculating the electron-phonon matrix elements in the proximity of the Fermi energy on 
fine grids of $\bf{k}$-points and $\bf{q}$-points of 64$\times$64$\times$32 and 32$\times$32$\times$16, 
respectively. The Fermi window used is 0.4 eV. We also examine the convergence of the gap function and 
anisotropic EPC strength with respect to both fine $\bf{k}$- and $\bf{q}$-mesh samplings, ensuring 
that the final result is well-converged.

\begin{figure}[!htb]
\centering
\includegraphics[width=0.90\columnwidth]{BiPdFig2.eps}
\caption{(a) Phonon dispersion, (b) phonon DOS (PhDOS), (c) Eliashberg spectral
function ($\alpha^{2}F$), and accumulative electron-phonon coupling constant ($\lambda(\omega)$)
of $\gamma$-BiPd.}
\label{fig:phonon_dis}
\end{figure}

\section{Crystal Structure and Phonon Dispersion}
\label{sec:Crystal_Structure}
The unit cell of $\gamma$-BiPd contains two formula units (two Bi and two Pd atoms per unit cell) organised in 
a hexagonal structure with space group P6$_3$/mmc (No. 194) (Fig.~\ref{fig:crystal}). 
In the present calculations, the fully optimised lattice constants $a=4.2719$ $\mathrm{\AA}$ 
and $c=5.6244$ $\mathrm{\AA}$ are used, which agree well with the experimental lattice parameters 
of $a=4.23$ $\mathrm{\AA}$ and $c=5.69$ $\mathrm{\AA}$ \cite{amcsd} (with errors within $\sim 1\%$).
The Wyckoff positions of Pd atoms and Bi atoms are 2a (0, 0, 0) and 2c (1/3, 2/3, 1/4), respectively (Fig.~\ref{fig:crystal}). 
Each Bi atom is coordinated by six Pd atoms, while each Pd atom is bonded to six Bi atoms and two additional Pd atoms.
The Bi-Pd bond length is 2.8391 $\mathrm{\AA}$, while the Pd-Pd bond length is slightly shorter 
at 2.8122 $\mathrm{\AA}$. The nearest Bi–Bi separation is 3.7406 $\mathrm{\AA}$. The Pd-Bi-Pd 
bond angle is $59.375^\circ$, and the Bi-Pd-Pd bond angle is $60.312^\circ$.

\begin{figure}[!htb]
\centering
\includegraphics[width=0.98\columnwidth]{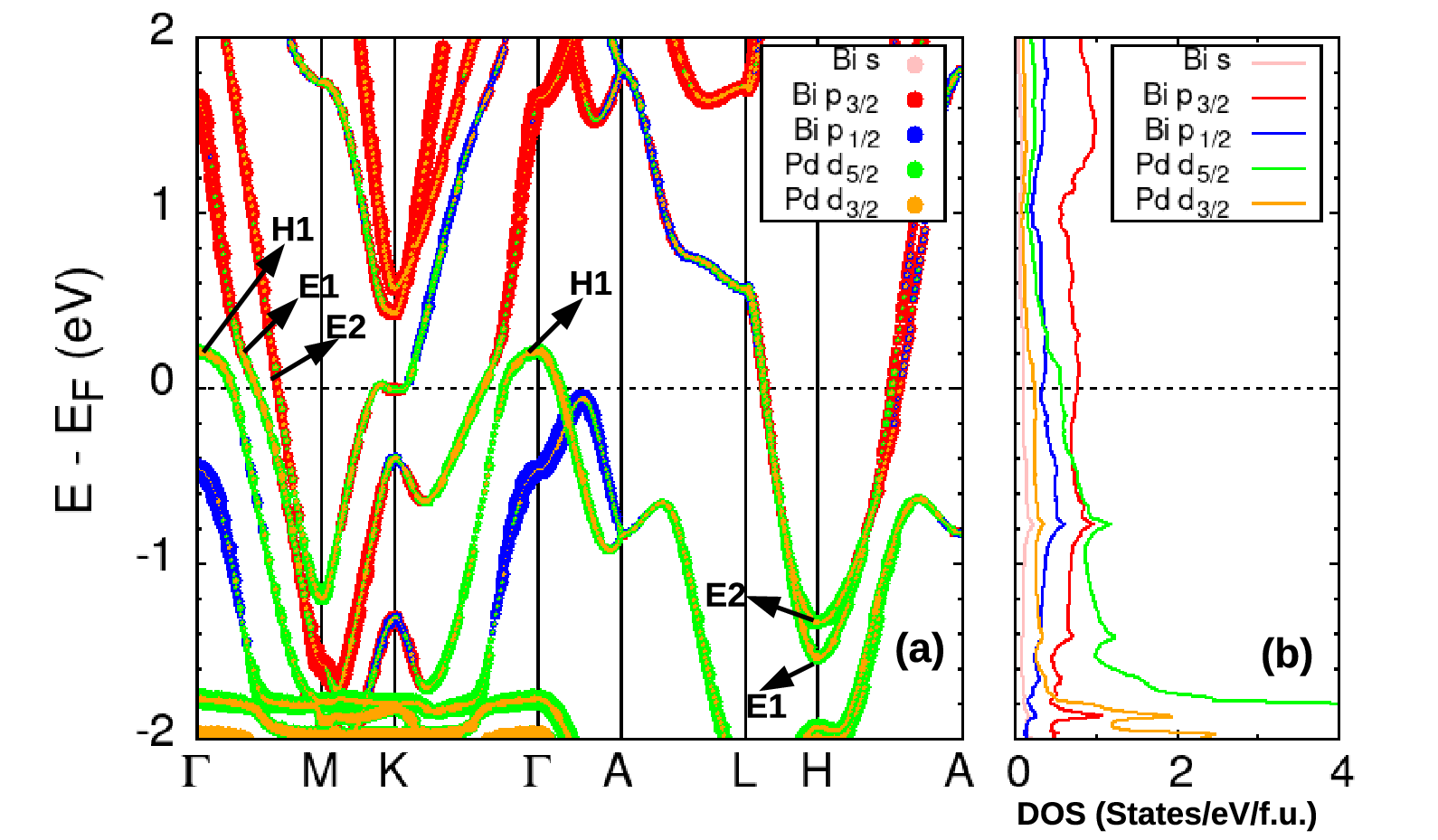}
\caption{(a) Orbital-projected fat plot of the band structure, with the orbital weights
being proportional to the sizes of the corresponding colored dots and (b) orbital-projected DOSs of $\gamma$-BiPd.
In (a), one band labelled H1 represents the hole pocket, and two bands labelled
E1 and E2 form the two electron pockets.}
\label{fig:bands_pdos}
\end{figure}

The calculated phonon dispersion and phonon DOS are presented in Figs.~\ref{fig:phonon_dis}(a) 
and \ref{fig:phonon_dis}(b), respectively. The phonon spectrum of $\gamma$-BiPd includes 12 modes: 
three acoustic and nine optical ones, as shown in Fig.~\ref{fig:phonon_dis}(a). Below 12.3 meV, 
the phonon DOS is dominated by Bi-atomic vibrations, attributed to the heavier mass of Bi atoms. 
Above this energy, Pd atomic vibrations become increasingly dominant. The $\Gamma$-point in the
Brillouin zone (BZ) exhibits centrosymmetric point group symmetry $D_{6h}(6/mmm)$. Within this symmetry, 
the longitudinal acoustic phonon mode corresponds to the $A_{2u}$ representation, 
while the two transverse acoustic modes belong to the $E_{1u}$ representation.
In the acoustic region, the transverse modes exhibit higher frequencies than the longitudinal mode. 
In contrast, in the optical region, the longitudinal optical modes have higher frequencies 
than the transverse optical modes.

\begin{figure*}[!htb]
\centering
\includegraphics[width=1.7\columnwidth]{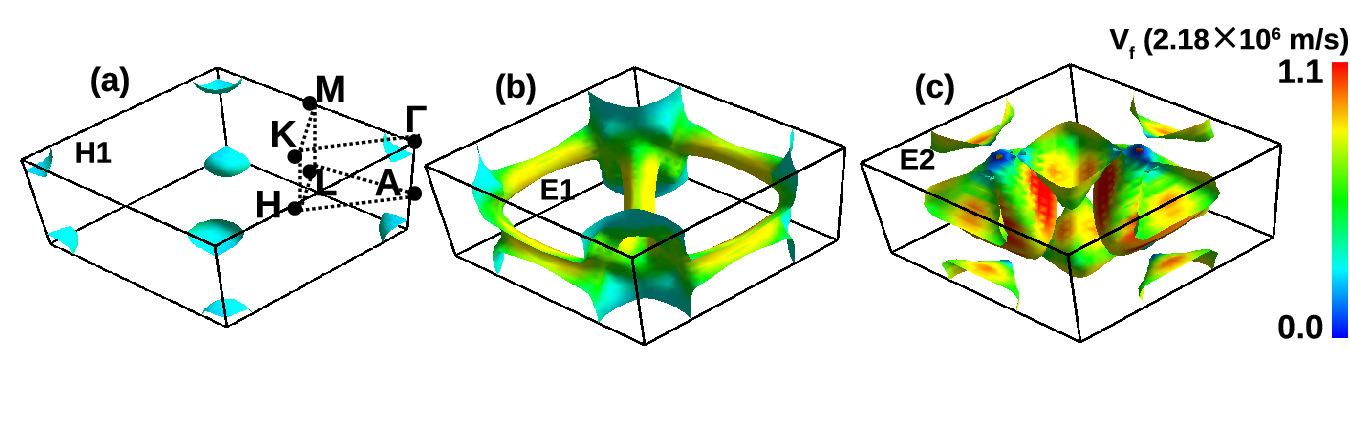}
\caption{Momentum ${\bf k}$-dependent magnitude of the Fermi velocity $v_f$ on the Fermi surface of $\gamma$-BiPd
on (a) hole pocket H1, (b) electron pocket E1, and (c) electron pocket E2.}
\label{fig:FS_vf}
\end{figure*}

\section{Electronic Structure and Fermi Surface}
\label{sec:Electronic_Structure}
In Fig. 3(a) and Fig. A1, we show fully relativistic electronic band structure of $\gamma$-BiPd,
and in Fig. 3(a), the sizes of Bi and Pd orbital contributions are indicated by the sizes of color dots.
In Fig. 3(b), Bi and Pd orbital-projected DOS spectra are displayed.  
In the presence of the SOC, atomic orbitals of orbital angular momentum $l$ would split into two sets of
sub-orbitals with total angular momentum $j = l-s$ and $j+s$, respectively. Here $s = 1/2$ is the spin angular momentum.
In particular, $p$-orbitals would split into doubly degenerate $p_{1/2}$-orbitals of $j=1/2$ and 
four-fold degenerate $p_{3/2}$-orbitals of $j=3/2$. And $d$-orbitals would split into four-fold 
degenerate $d_{3/2}$-orbitals of $j=3/2$ and six-fold degenerate $d_{5/2}$-orbitals of $j=5/2$.  
In the vicinity of the Fermi level, the DOS is mainly of Bi $p$-orbital and Pd $d$-orbital characters 
[Fig. 3(b)]. For example, Bi $p_{3/2}$-orbitals dominates the DOS near the $E_F$,
while the second largest contribution comes from Pd $d_{5/2}$-orbitals. 
Specifically, total DOS at the $E_F$ $N_F = 1.10$ states/eV/BiPd, and atom-decomposed DOSs at the $E_F$
are 0.644 and 0.416 states/eV/atom, respectively, for Bi and Pd atoms. Orbital-projected DOSs at the $E_F$
for Bi $s$, $p_{1/2}$ and $p_{3/2}$ as well as Pd $d_{3/2}$ and $d_{5/2}$ are, respectively,
0.054, 0.180, 0.409, 0.119 and 0.294 states/eV/atom. Interestingly, Fig. 3(b) indicates
that all the orbital-projected DOS spectra are rather flat in the vicinity of the $E_F$
and are almost constant over a narrow energy window of 0.1 eV around the Fermi level. 

Figures 3(a) and A1 show that three bands [labelled H1, E1 and E2 in Fig. 3(a)] cross the $E_F$, indicating metallic behavior 
of $\gamma$-BiPd. As a result, the Fermi surface consists of three Fermi sheets (pockets), as shown in Fig. ~\ref{fig:FS_vf}. 
The H1 band forms a hole pocket centered at the $\Gamma$ point, while the E1 and E2 bands form two large electron-like Fermi sheets. 
Analyzing the orbital contributions near the high-symmetry points [Fig.~\ref{fig:bands_pdos}(a)], 
we find that the H1 band near the $E_F$ at the $\Gamma$-point arises primarily from Pd $d$-orbitals. 
Their relatively localized nature results in a low Fermi velocity $v_f$ [Fig.~\ref{fig:FS_vf}(a)].
On the other hand, the E1 and E2 electron pockets span a large portion of the BZ [Figs.~\ref{fig:FS_vf}(b) and (c)], 
and thus contribute more significantly to the DOS near the $E_F$. 
These bands are highly dispersive and anisotropic, 
leading to larger and direction-dependent Fermi velocities. For instance, the E2 band at the K point 
near the $E_F$ exhibits strong hybridization between Pd $d_{3/2}$- and $d_{5/2}$-orbitals 
and Bi $p_{3/2}$-orbitals. It forms an open neck at the K-point with a relatively low $v_f$, 
while along the $\Gamma-$A direction, its dispersive character results in a high $v_f$.

\section{Electron-phonon Coupling}
\label{sec:EPC}
For phonon-mediated superconductors, the ME theory has proven effective in predicting both the $T_c$ and the anisotropic 
SC gap \cite{Anisotropic_Margine2013,Origin_Heil2017}. Two main ingredients, namely, Eliashberg spectral function $\alpha^{2}F$
and effective Coulomb repulsion potential $\mu^{*}_{c}$, are required 
to solve the ME equations [Eqs.~(\ref{eq:1}) and (\ref{eq:2})] \cite{Anisotropic_Margine2013}. 
The effective Coulomb repulsion enters the formalism via the momentum-dependent Coulomb matrix elements 
$V({\bf{k-k^{'}}})$ [Eq. (\ref{eq:2})] \cite{Electron_lee2023}. The Coulomb potential is double-averaged over the FS, 
yielding an effective Coulomb interaction $\mu_{c}=N_{F}\langle\langle V({\bf{k-k^{'}}})\rangle\rangle_\mathrm{FS}$. 
In the ME equations, this averaged interaction replaces the term $N_{F}V({\bf{k-k^{'}}})$ with its renormalized form $\mu_{c}^{*}$,
called Morel-Anderson pseudopotential. This Coulomb pseudopotential $\mu_{c}^{*}$ has often been treated as a semiempirical 
parameter \cite{EPW_ponce2016,Electron_lee2023,EPW_noffsinger2010}, and the $\mu_{c}^{*}$ value of $\sim$0.10 has been widely
used for conventional superconductors especially classic elemental superconductors~\cite{Eliashberg_Pellegrini2022}.
It has also been found that $\mu_{c}^{*} \approx 0.05\sim0.10$ works quite well for many $sp$-electron systems,
while $\mu_{c}^{*} \approx 0.10\sim0.15$ is quite appropriate for transition metal compounds~\cite{Eliashberg_Pellegrini2022}.

\begin{figure}[!htb]
\centering
\includegraphics[width=0.80\columnwidth]{BiPdFig5.eps}
\caption{Total and band-resolved normalized density distribution of (a) the EPC strength $\lambda_{n{\bf{k}}}$,
$g(\lambda_{n{\bf k}})$, and of (b) SC gap $\Delta_{n{\bf{k}}}$, $g(\Delta_{n{\bf k}})$, on the FS of $\gamma$-BiPd.}
\label{fig:lambda_gap}
\end{figure}

The Eliashberg spectral function $\alpha^{2}F$, shown in Fig.~\ref{fig:phonon_dis}(c), represents the frequency-resolved EPC. 
The shape of $\alpha^{2}F$ closely follows the phonon DOS, consisting of three distinct peaks: 
one corresponding to the acoustic phonons and two arising from the optical phonons.
The phonon DOS spectrum can be divided into two regions at energy of 12.3 meV. 
This separation originates from the significant mass difference between Bi and Pd atoms. This separation is followed
up in $\alpha^{2}F$, reflecting the atomic character of EPC across the frequency spectrum. Specifically, 
the portion of $\alpha^{2}F$ below 12.3 meV, including two prominent peaks, is primarily attributed to Bi-related vibrational modes. 
In contrast, the broader peak above 12.3 meV is predominantly governed by the vibrations of lighter Pd atoms.

The total EPC constant $\lambda$, also known as the mass enhancement parameter, is obtained by integrating $\alpha^{2}F$/$\omega$
over the entire frequency range. The calculated EPC constant $\lambda$ of $\gamma$-BiPd is 0.46, and
this places $\gamma$-BiPd among the intermediate strong-coupled superconductors such as Al and Ga~\cite{Transition_McMillan1968}. 
From the cumulative $\lambda$ shown in Fig.~\ref{fig:phonon_dis} (c), we observe that
approximately 80$\%$ of the total $\lambda$ arises from Bi-atomic vibrations occurring below 12.3 meV.
The remaining contribution comes from Pd-atomic vibrations. This clearly highlights the dominant role of Bi atoms in the EPC, 
particularly within the acoustic phonon region. 
Since the phonon DOS spectrum of $\gamma$-BiPd could be roughly separated into the Bi vibration dominent low frequency region
and the Pd oscilation dominent high frequency region at 12.3 meV [(Fig. 2(b)], 
the EPC strength $\lambda$ could be broken down into contributions from the electronic stiffness $K_{e,j}$ and the lattice 
stiffness $K_{l,j}$ of the two atoms ($j$ denotes either Bi or Pd), expressed 
as $\lambda_{j}=\frac{K_{e,j}}{K_{l,j}}$ \cite{Room_Pickett2023}.
The $K_{e,j}$ and $K_{l,j}$ are known to be proportional to the atom-decomposed DOS at the $E_F$
and the second moment of phonon frequency, respectively~\cite{Room_Pickett2023}. As discussed above, in $\gamma$-BiPd, 
Bi atom would has a larger DOS at the $E_F$ than Pd atom, and would have a lower second moment of phonon frequency
than Pd. Consequently, Bi atom would have a higher $K_e$ and a lower $K_l$ than Pd atom,
thus resulting in a larger contribution to the EPC constant $\lambda$ than Pd atom.

\begin{figure*}[!htb]
\centering
\includegraphics[width=1.7\columnwidth]{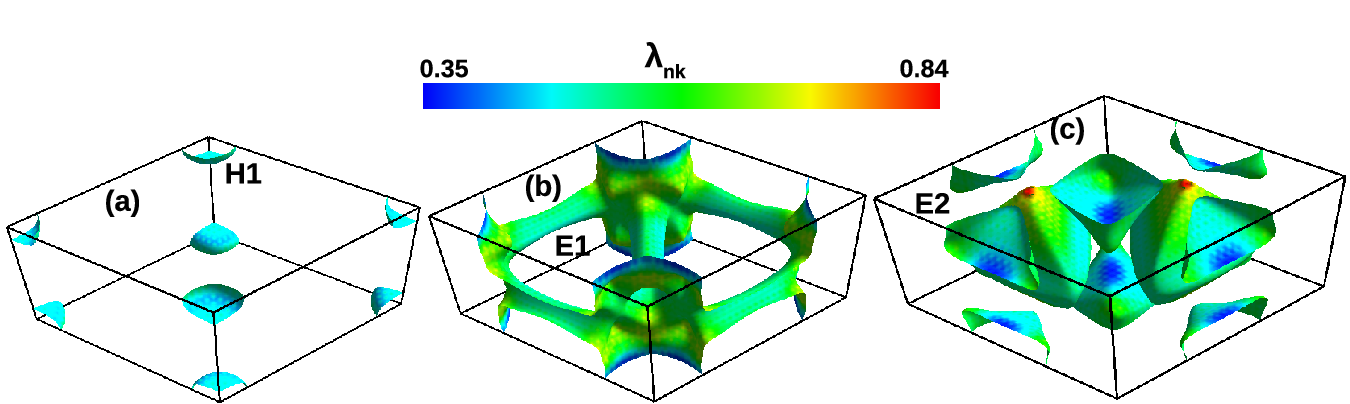}
\caption{Momentum ${\bf k}$-dependent EPC strength $\lambda_{\bf{k}}$ on the Fermi surface of $\gamma$-BiPd
on (a) hole pocket H1, (b) electron pocket E1, and (c) electron pocket E2.}
\label{fig:lamnda_FS}
\end{figure*}

In Fig.~\ref{fig:lambda_gap}(a), we present the normalized distribution of total and band-decomposed EPC strength 
$\lambda_{{\bf{k}}}$ on the FS. This quantity represents the probability density of the electron-phonon interaction 
over the FS. The distribution is centered around $\lambda_{\bf k}=0.46$ with a rather broad width, 
indicating significant anisotropy in the EPC on the FS. Figure~\ref{fig:lambda_gap}(a) indicates that
bands E1 and E2 make dominent contributions to the EPC.  

\begin{figure*}[!htb]
\centering
\includegraphics[width=1.7\columnwidth]{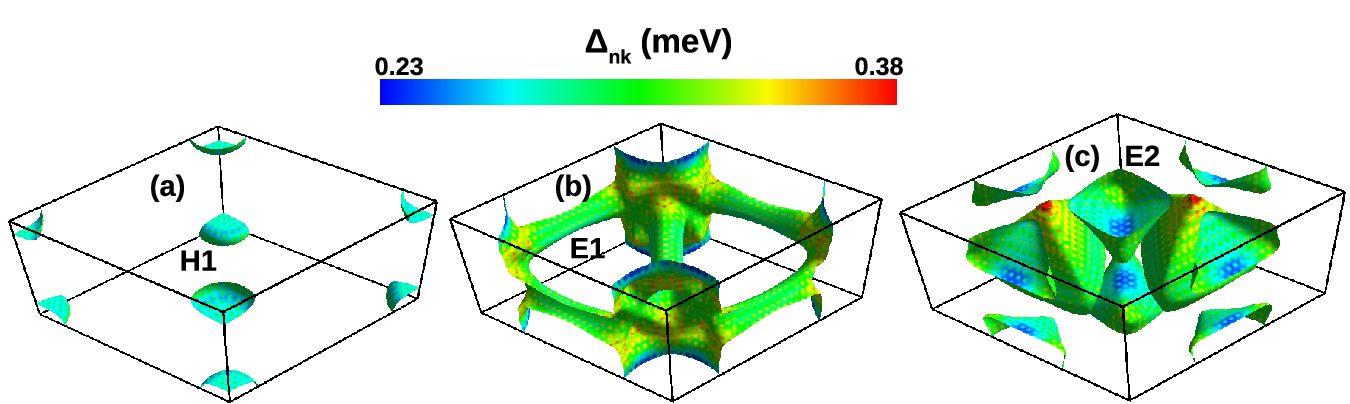}
\caption{Momentum ${\bf{k}}$-dependent SC gap $\Delta_{n{\bf k}}$ on the Fermi surface of $\gamma$-BiPd at 0.5 K
on (a) hole pocket H1, (b) electron pocket E1, and (c) electron pocket E2.}
\label{fig:gap_FS}
\end{figure*}

Figures~\ref{fig:lamnda_FS} show momentum-resolved $\lambda_{\bf{k}}$ on the three Fermi sheets.
The $\lambda_{{\bf{k}}}$ would contribute to 
the renormalization of both the energy bands and the DOS at the $E_{F}$, quantified by the renormalization function 
$Z_{{\bf{k}}}=1+\lambda_{{\bf{k}}}$. Fermi velocity $v_{{\bf{k}}}$ would be reduced by a factor of $Z_{{\bf{k}}}$, 
while the DOS and related quantities such as electronic specific heat would be enhanced by the same factor. 
In the parabolic approximation, the effective mass $m^{*}$ is related with $v_{{\bf{k}}}$ 
as $m^{*}=\hbar k_{F}/v_{{\bf{k}}}$, leading to an enhancement of $Z_{\bf k}$.
These renormalizations due to the EPC can be measured in thermodynamic experiments such as specific heat 
and thermal conductivity. For example, the electronic contribution to the specific heat of a free-electron
metal in the low temperature limit is given by $C_{e}=\gamma_n T$ where normal state Sommerfeld 
coefficient $\gamma_n = \frac{\pi^2}{3}k^{2}_{B}N_{F}$. In the presence of the EPC, $\gamma_n$ would be
enhanced by a factor of (1.0 + $\lambda$) because $N_F$ would be increased by the same factor.
For $\gamma$-BiPd, $N_F = 1.10$ states/eV/BiPd and $\lambda = 0.46$. Therefore, the $\gamma_n$ would 
be $0.097$ mJ/K$^2$cm$^3$ and $0.141$ mJ/K$^2$cm$^3$, respectively, before and after
considering the EPC enhancement. The experimental $\gamma_n$ value has not been reported
for $\gamma$-BiPd. Nevertheless, specific heat experiments were recently performed on
the Bi(Pd$_{0.5}$Pt$_{0.5}$) alloy, which is both isoelectronic and isostructural 
to $\gamma$-BiPd~\cite{Sharma2024}.
The measured $\gamma_n$ for Bi(Pd$_{0.5}$Pt$_{0.5}$) is $0.119$ mJ/K$^2$cm$^3$, which is
close to our predicted EPC-enhanced $\gamma_n$ value. 

\section{Superconducting Properties}
\label{sec:Superconducting_Properties}
To obtain the SC gap function $\Delta_{n\mathbf{k}}$, we solve the anisotropic ME equations at several 
temperatures using $\mu_c^{*} = 0.05$~\cite{Eliashberg_Pellegrini2022,Li2024}, which is adopted here
mainly because the DOS near the $E_F$ is dominated by Bi $p$-orbitals (see Fig. 3). The influence 
of the choice of $\mu_c^{*}$ on the calculated SC gap function and $T_c$ is discussed in Appendix~\ref{sec:appenix_B} 
(Also see, e.g., Refs.~\cite{Eliashberg_Pellegrini2022} and \cite{Li2024}). 
In Fig.~\ref{fig:lambda_gap}(b), we display the normalized distribution of total and band-decomposed SC gap $\Delta_{n\bf{k}}$ 
calculated at 0.6 K. The distribution is clustered around $\Delta_{\bf k}=0.31$ meV with a rather broad width of 0.15 meV,
suggesting significant anisotropy in the SC gap on the FS. Figure~\ref{fig:gap_FS} shows momentum-resolved SC 
gap $\Delta_{n\bf{k}}$ at 0.5 K on the FS, and indicates that the gap opening occurs on all three Fermi sheets. 
The $\Delta_{n\bf{k}}$ exhibits pronounced anisotropy across the FS. Electron Fermi sheet E2 exhibits 
the largest anisotropy [Fig. \ref{fig:gap_FS}(c)], while hole pocket H1 displays the least anisotropy [Fig. \ref{fig:gap_FS}(a)]. 
The calculated lowest and highest gap magnitudes are 0.23 meV and 0.38 meV, respectively, located on the E2 Fermi sheet.
In particular, the largest SC gap is located near the K-point on a narrow neck of the E2 Fermi sheet [Fig. \ref{fig:gap_FS}(c)].

Figure~\ref{fig:lambda_gap}(b) shows that $\gamma$-BiPd is a single-gapped anisotropic superconductor,
although the $\Delta_{n\bf{k}}$ varies across the FS (Figure~\ref{fig:gap_FS}), as mentioned above.
This is consistent with the spin-singlet $s$-wave superconductivity observed in recent quantum oscillation measurements 
on $\gamma$-BiPd films \cite{Unequivocal_Chiang2023}.
Interestingly, recent experiments (see, e.g.,~\cite{Herrera2015,Single_Kacmarcik2016}) indicated that $\beta$-Bi$_2$Pd also has  
a single $s$-wave superconducting gap, although earlier experiments~\cite{Superconductivity_imai2012}
suggested it to be a multigap superconductor based on the anomalous temperature dependences of
the upper critical magnetic field and specific heat measurements. 
From Fig.~\ref{fig:lambda_gap}, Fig. ~\ref{fig:lamnda_FS} and Fig. \ref{fig:gap_FS}, a clear correspondence 
between $\lambda_{\bf{k}}$ and $\Delta_{n\bf{k}}$ is evident.
This could be expected because $\gamma$-BiPd is predominently a phonon-mediated superconductor. 
This behavior has been also observed in other conventional superconductors such as NbS$_2$ \cite{Origin_Heil2017} 
and PbTaSe$_2$ \cite{Lian2019}.

The anisotropy in the gap function of $\gamma$-BiPd can be further understood from the orbital characterization of 
the energy bands near the Fermi level [Fig.~\ref{fig:bands_pdos}(a)]. The SC gap on the H1 Fermi sheet primarily originates 
from Pd $d$-orbitals. In contrast, the E1 and E2 Fermi sheets exhibit predominant Bi $p$-orbital characters. 
Since the E1 and E2 Fermi sheets span a large portion of the BZ and dominate
the DOS near the $E_F$, the SC gap in $\gamma$-BiPd is largely governed by Bi $p$-orbitals.
This Bi dominance is further corroborated by the fact that the EPC strength is mainly derived 
from the acoustic and low-energy optical phonon modes, which are predominantly associated with Bi atomic vibrations
[see Figs.~\ref{fig:phonon_dis}(b) and \ref{fig:phonon_dis}(c)].
Nonetheless, the role of Pd $d$-orbitals, particularly Pd $d_{5/2}$ components, remains non-negligible, 
especially in shaping the SC properties of the H1 band. The K-point which
exhibits the largest SC gap magnitude on the E2 Fermi sheet is contributed mainly from
hybridization of Pd $d$ and Bi $p_{3/2}$-orbitals.
These orbital-specific participations suggest that the superconductivity in $\gamma$-BiPd is inherently orbital-selective, 
with the pairing amplitude modulated by the orbital-projected DOS involved. This provides a microscopic basis 
for the anisotropic yet single-gap superconductivity found in the system.

Such orbital-selective behaviors have been found in other SC materials, including  iron-based superconductors 
FeSe \cite{Discovery_sprau2017,Orbital_nica2017,Orbital_Kreisel2017,Imaging_kostin2018,Orbital_hu2018} 
and LaFeAsO$_{1-x}$F$_x$ \cite{Unconventional_Kuroki2008,Unconventional_Mazin2008,Even_Dai2008,Role_yi2017,Multiorbital_nica2021}
as well as conventional superconductors MgB$_2$, NbSe$_2$ \cite{Fermi_yokoya2001} and NbS$_2$ \cite{Origin_Heil2017,Orbital_Bi2022}. 
In particular, three Fe $d_{xz}$, $d_{yz}$ and $d_{xy}$ orbitals dominate the electronic states near the $E_F$ 
and thus give rise to the multiorbital superconductivity in FeSe. 
In LaFeAsO$_{1-x}$F$_x$, although there is significant hybridization among five Fe $d$-orbitals, 
which form four Fermi sheets, the SC gap retains the $s$-wave symmetry with opposite signs between electron 
and hole pockets. Similarly, in conventional superconductor 2H-NbS$_2$ \cite{Origin_Heil2017}, 
three Fermi sheets are present, and pairing contributions from S $p_z$ and Nb $d$-orbitals 
exhibit different amplitudes. It leads to two distinct and ${\bf{k}}$-dependent SC gaps 
as measured experimentally \cite{Superconducting_Guillamon2008} and latter explained 
theoretically \cite{Origin_Heil2017}. The anisotropic and orbital dependent SC gap 
can be measured using quasiparticle interference imaging \cite{Superconducting_Guillamon2008}, 
high resolution angle-resolved photoemission spectroscopy (ARPES) \cite{Angle_Damascelli2003,Distinct_Mou2011}, 
and scanning tunneling microscopy and spectroscopy (STM and STS) \cite{Scanning_Fischer2007}.

\begin{figure}[!htb]
\centering
\includegraphics[width=0.90\columnwidth]{BiPdFig8.eps}
\caption{(a) SC gap $\Delta$ as a function of temperature $T$, with the red dots denoting
the averaged gap values, which are fitted using a BCS-type temperature dependence [Eq. (7)]
(the dashed red curve).
(b) Normalized SC quasiparticle DOS $N_s/N_F$ as a function of energy calculated at several temperatures.}
\label{fig:gap-T-Ns}
\end{figure}

In Fig.~\ref{fig:gap-T-Ns}(a), the SC gap $\Delta$ calculated  
at several different temperatures are plotted. To evaluate the superconducting
$T_c$, the averaged SC gap values at the considered temperatures [denoted by red dots 
in Fig.~\ref{fig:gap-T-Ns}(a)] 
are fitted using a BCS-type temperature dependence~\cite{Microscopic_BCS1957} 
\begin{align} 
\Delta(T) = \Delta(0)\mathrm{tanh(\alpha\sqrt{T_c/T-1})}.
\end{align}
We obtain $\Delta(0)=0.305$ meV, $\alpha = 1.90$ and $T_c = 1.96$ K. 
The calculated $T_c$ of 1.96 K agree reasonably well with the experimental value of $\sim$3.3 K reported
for $\gamma$-BiPd thin films \cite{Unequivocal_Chiang2023}. 
The obtained $\alpha$ of 1.90 is slightly larger than 1.73 from the BCS theory, and this deviation
from the universal BCS value may be attributed to the participation of multibands and SC gap anisotropy.
With the knowledge of $v_f$ (Fig.~\ref{fig:FS_vf}) and SC gap, 
the coherence length $\xi$ can be estimated using the BCS relation $\xi=\hbar v_{f}/\pi \Delta$. 
The calculated average values of $v_{f}$ and $\Delta$ 
are $\sim$1.2 $\times 10^{6}$ m/s and $\sim$0.305 meV, respectively,
giving rise to $\xi$ of 0.72 $\mu$m. Interestingly, this $\xi$ is about ten times larger than
that of $\alpha$-BiPd (66 nm)~\cite{Dirac_sun2015}, and may be attributed to
the higher Fermi velocity and lower SC gap magnitude in $\gamma$-BiPd. 

Using Eq.~(\ref{eq:3}), we also calculate the normalized quasiparticle DOS $\frac{N_s}{N_F}$
as a function of energy at several temperatures for $\gamma$-BiPd, as displayed in Fig.~\ref{fig:gap-T-Ns}(b). 
Clearly, there is only one peak on the SC gap edge, indicating a single gap superconductivity
in $\gamma$-BiPd. Furthermore, the $\frac{N_s}{N_F}$ spectrum shows a U-shaped function,
further indicating a spin-singlet $s$-save superconductivity in $\gamma$-BiPd (see, e.g., ~\cite{Herrera2015,Li2024}).   
This predicted $\frac{N_s}{N_F}$ spectra can be measured by using, e.g., scanning tunneling microscopy
and spectroscopy (STM and STS)\cite{Scanning_Fischer2007}. Indeed, this has recently been done for $\beta$-Bi$_2$Pd~\cite{Herrera2015}.

\section{Conclusion}
\label{sec:Conclusion}
In summary, we have studied the SC properties of $\gamma$-BiPd by solving the anisotropic Migdal-Eliashberg equations 
in conjunction with {\it ab initio} relativistic calculations of the electron and phonon band structures as well as EPC matrix elements.
We find that $\gamma$-BiPd possesses a complex FS, consisting of two electron pockets and one hole pocket, 
each characterised by distinct atomic orbitals. Importantly, we discover that the superconductivity in 
$\gamma$-BiPd is orbital-selective, arising from Bi $p$-orbitals, and distributed anisotropically on the FS.
While our results show momentum {\bf k}-dependent EPC strength $\lambda_{\bf k}$ and SC gap $\Delta_{\bf k}$
across the FS, that calculated superconducting quasiparticle density of states $N_S$ spectra exhibit a U-shaped gap 
and $\Delta_{\bf k}$ distribution forms a single peak, indicates a single gap spin-singlet $s$-wave superconductivity 
in this material. The calculated $T_c$ is $\sim$2.0 K, agreeing in order of magnitude with the experimental value 
of 3.3 K in $\gamma$-BiPd thin films~\cite{Unequivocal_Chiang2023}. Interestingly, the evaluated coherence length $\xi$ of 0.72 $\mu$m is 
about ten times larger than that of $\alpha$-BiPd (66 nm)~\cite{Dirac_sun2015}.
The predicted EPC-enhanced Sommerfeld coefficient $\gamma_n$ of $0.141$ mJ/K$^2$cm$^3$ 
is close to the experimental $\gamma_n$ value ($0.119$ mJ/K$^2$cm$^3$) of the isoelectronic 
and isostructural Bi(Pd$_{0.5}$Pt$_{0.5}$) alloy~\cite{Sharma2024}. 
Our interesting findings will stimulate further experiments on $\gamma$-BiPd,
such as scanning tunneling microscopy and spectroscopy, quasiparticle interference imaging,
high resolution angle-resolved photoemission spectroscopy and low temperature specific heat experiments. 


\section*{Acknowledgments}
The authors acknowledge the support from the National Science and Technology Council (NSTC) and National Center for
Theoretical Sciences (NCTS) in Taiwan. The authors also thank the National Center for High-performance Computing (NCHC)
in Taiwan for the computing time.

\appendix
\renewcommand{\thefigure}{A\arabic{figure}}
\setcounter{figure}{0}
\section{Effect of Spin-Orbit Coupling on the Electronic Band Structure}
\label{sec:appenix}
\begin{figure}[h!]
\centering
\includegraphics[width=0.90\columnwidth]{BiPdFig9.eps}
\caption{Band structures of $\gamma$-BiPd calculated without (blue curves) and with (red curves) the SOC included.}
\label{fig:bands_nosoc_soc}
\end{figure}
Due to the presence of heavy Bi atoms in $\gamma$-BiPd, the relativistic SOC is expected to change the electronic
band structure significantly. Thus, to examine the SOC effects, we have calculated the band structure
of $\gamma$-BiPd both with and without including the SOC, and the results are shown in Fig. A1. 
It is clear from Fig. A1 that the presence of the SOC substantially alters the band structure, especially
the energy bands near the $E_F$. In particular, the SOC opens a large gap of about 1.5 eV at the Dirac nodal
point at $\sim$1.4 eV above the $E_F$ along the $\Gamma$-A symmetry line. This greatly pushes the lower band 
down to below the Fermi level, resulting in the disappearance of the large hole pocket centered at the $\Gamma$ point. 
Figure A1 also shows that the presence of the SOC greatly changes the band structure and FS topology near the K point
in the vicinity of the $E_F$. Therefore, because of these significant effects of the SOC, 
we present only the results from the fully relativistic calculations in this paper. 

\section{Effect of Coulomb pseudopotential $\mu^{*}_c$ on SC gap function and $T_{c}$}
\label{sec:appenix_B}
\begin{table}[b]
\caption{\label{tab:A1}%
The average SC gap $\Delta_{avg}$ on the Fermi surface and transition temperature $T_{c}$ for four 
different $\mu^{*}_{c}$ values. $T_{c}$ is estimated using the ADM formula [Eq. (B1)] 
as well as the calculated EPC constant $\lambda=0.46$ and $\hbar\omega_{log}=8.3356$ meV. }
\begin{ruledtabular}
\begin{tabular}{lcdrr}
$\mu^{*}_{c}$ & 0.03 & 0.05 & 0.07 & 0.09\\
$\Delta_{avg}$ & 0.373 & 0.303 & 0.243 & 0.193\\
$T_{c}$ & 2.196 & 1.738 & 1.331 & 0.980\\
\end{tabular}
\end{ruledtabular}
\end{table}

\begin{figure}[h!]
\centering
\includegraphics[width=0.98\columnwidth]{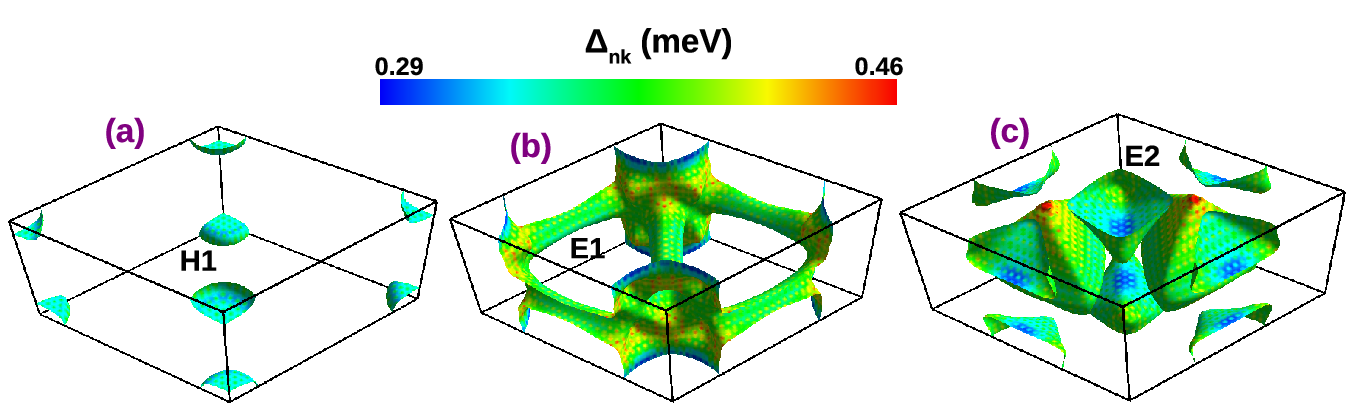}
\includegraphics[width=0.98\columnwidth]{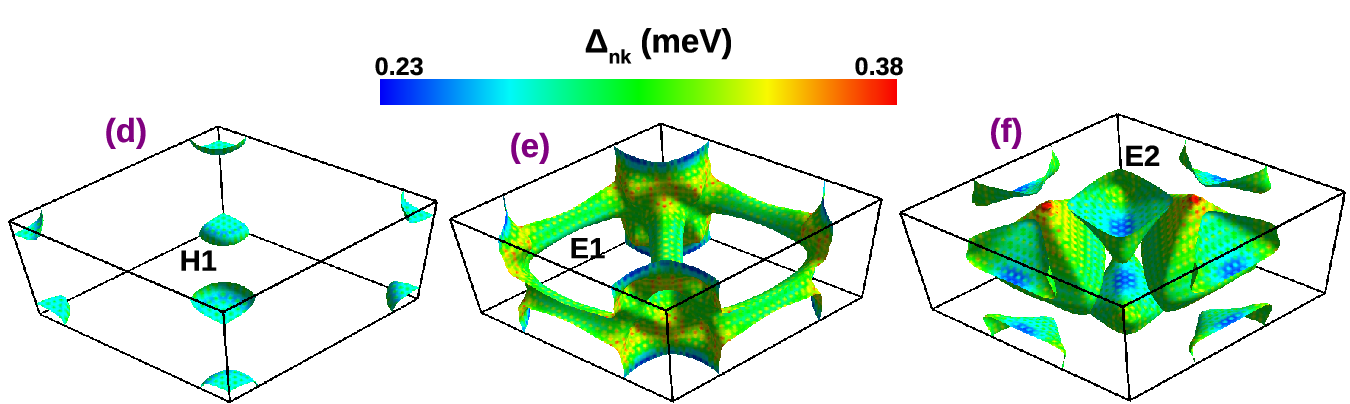}
\includegraphics[width=0.98\columnwidth]{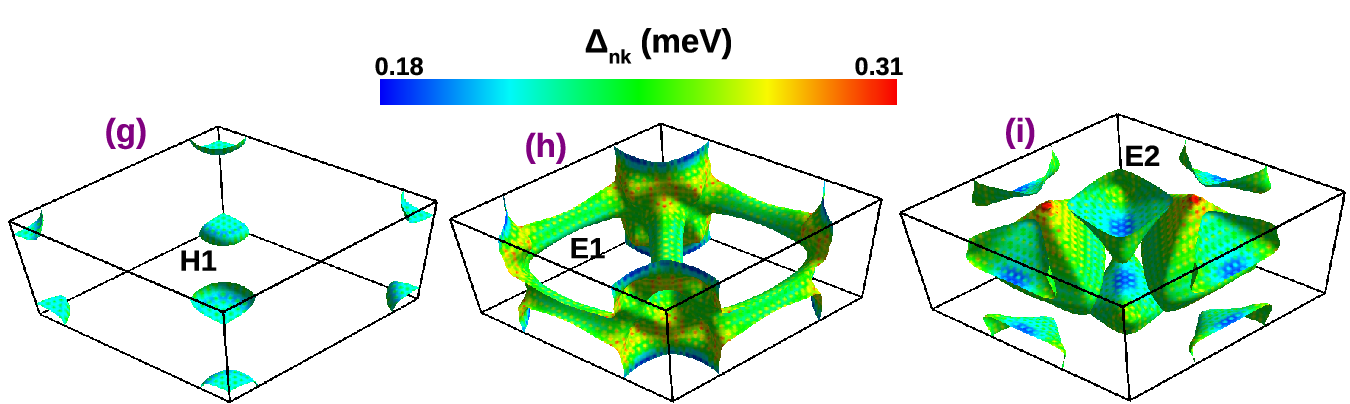}
\includegraphics[width=0.98\columnwidth]{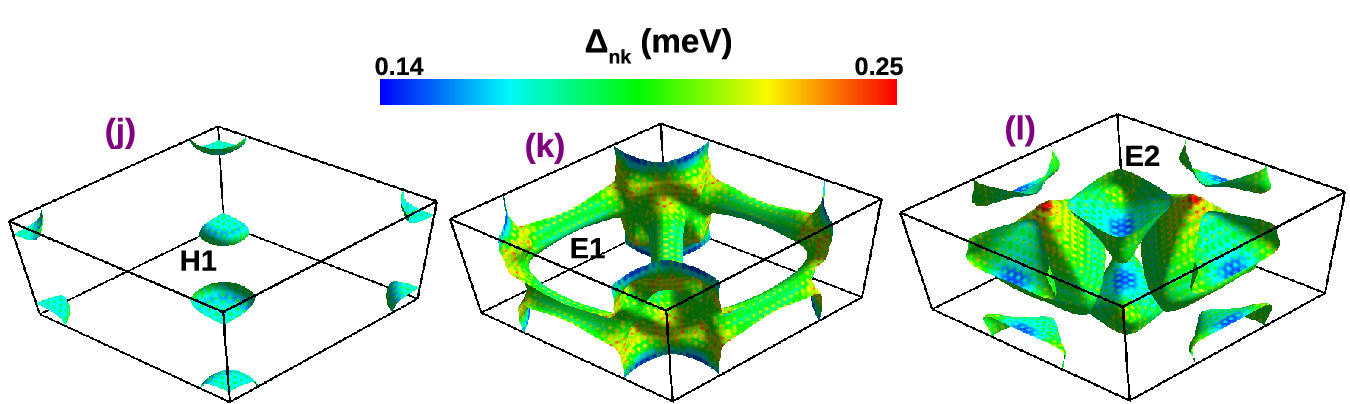}
\caption{Wave-vector {\bf k}-dependent SC gap function $\Delta_{n{\bf k}}$ at 0.6 K on the Fermi surface
of $\gamma$-BiPd, calculated using $\mu^{*}_c$ of (a)--(c) 0.03, (d)--(f) 0.05, (g)--(i) 0.07, and (j)--(l) 0.09.
}
\label{fig:gap_VS_mu}
\end{figure}

The value of Coulomb pseudopotential $\mu^{*}_c$ would affect the calculated SC $T_{c}$
and gap function $\Delta_{n\mathbf{k}}$. Generally, the larger $\mu^{*}_c$ value would
result in the lower $T_{c}$ and smaller $\Delta_{n\mathbf{k}}$.
For a specific $\mu^{*}_c$ value, solving the anisotropic ME equations [Eqs. (1-2)] for $\Delta_{n\mathbf{k}}$
for many different temperatures to determine the $T_{c}$, is computationally extremely demanding, especially
for the small SC gap cases~\cite{Li2024} and also when $T$ approaches to $T_{c}$. 
Thus, to examine the $\mu^{*}_c$-dependence of $T_{c}$ of $\gamma$-BiPd, we use the simplied 
Allen-Dynes McMillan (ADM) formula~\cite{Electron_lee2023} 
\begin{equation}
k_{B}T_{c}=\frac{\hbar \omega_{log}}{1.20}\mathlarger{\mathrm{exp\mathlarger{\mathlarger{[}}}\frac{-1.04(1+\lambda)}{\lambda-\mu^{*}_{c}(1+0.62\lambda)}\mathlarger{\mathlarger{\mathlarger{]}}}},
\label{eq:A1}
\end{equation}
together with the calculated EPC constant $\lambda$ and logarithmically averaged phonon frequency $\hbar\omega_{log}$
to estimate $T_c$, as listed in Table I. Nevertheless, we have also solved anisotropic ME equations [Eqs. (1-2)] 
at $T = 0.6$ K for four different $\mu^{*}_c$ values. The resultant SC gap functions $\Delta_{n\mathbf{k}}$ are 
displayed in Fig.~\ref{fig:gap_VS_mu} and the corresponding average gap $\Delta_{avg}$ values are listed in Table I.

As expected, Table I shows that $T_{c}$ decreases monotonically as $\mu^{*}_c$ increases. In particular, 
varying $\mu^{*}_c$ from 0.05 to 0.09 would reduce $T_{c}$ by 0.76 K, from 2.20 K to 0.98 K.
We note that $T_c$ determined by solving the ME equations for the $\Delta_{n\mathbf{k}}$ as a function of $T$ 
is slightly larger than that estimated using the ADM formula [Eq. (B1)]. For $\mu^{*}_c = 0.05$, $T_{c}$ 
from the solutions of the ME equations is 1.96 K, being larger than that (1.74 K) from the ADM formula (see Table I).
Figure~\ref{fig:gap_VS_mu} indicates that when the $\mu^{*}_c$ is altered, the pattern of SC gap function $\Delta_{n{\bf k}}$ 
hardly changes, although the magnitude of the SC gap decreases with increasing $\mu^{*}_c$ (see also Table I).
Our choice of the lower $\mu^{*}_c$ value of 0.05 is based on the following two considerations.
First, the theoretical $T_{c}$ value (1.98 K) is in order of magnitude agreement with the experimental $T_{c}$ value 
of $\sim$3.40 K~\cite{Unequivocal_Chiang2023}. Second, the magnitude of the SC gap function $\Delta_{n\mathbf{k}}$
is sufficently large so that stable solutions of the anisotropic ME equations could be numerically obtained
within the available computing resources. The main subjects of the present paper are the anisotropic nature and orbital 
character of the SC gap function $\Delta_{n{\bf k}}$. Since the pattern of the calculated SC gap function $\Delta_{n{\bf k}}$
remains rather robust again a reasonable variation of $\mu^{*}_c$ used, the main conclusions made here 
are independent of the specific $\mu^{*}_c$ value used in this work. Of course, treating $\mu^{*}_c$ as
a semiempircal parameter in the powerful ME theory is unsatisfactory. Indeed, fully {\it ab initio} ME theory 
in which the effective electron-electron Coulomb repulsion is determined from first-principles, 
has recently been developed~\cite{Eliashberg_Pellegrini2022}. Nevertheless, such fully {\it ab initio} ME theory calculations
are beyond the scope of the present paper. 
 

{}	

\end{document}